%% file: Drawdash.tex
\documentclass[letterpaper]{article} 
\usepackage{aaai2026}  
\usepackage{times}  
\usepackage{helvet}  
\usepackage{courier}  
\usepackage[hyphens]{url}  
\usepackage{graphicx} 
\usepackage{natbib}  
\usepackage{caption} 
\usepackage{algorithm}
\usepackage{algorithmic}

\usepackage{newfloat}
\usepackage{listings}
\DeclareCaptionStyle{ruled}{labelfont=normalfont,labelsep=colon,strut=off} 
\floatstyle{ruled}
\newfloat{listing}{tb}{lst}{}
\floatname{listing}{Listing}
\nocopyright

\title{Proactive Agentic Whiteboards: Enhancing Diagrammatic Learning}
\author{
    Suveen Ellawela\textsuperscript{\rm 1},
    Sashenka Gamage\textsuperscript{\rm 2},
    Dinithi Dissanayake\textsuperscript{\rm 1}
}
\affiliations{
    \textsuperscript{\rm 1}School of Computing, National University of Singapore, Singapore.\\
    \textsuperscript{\rm 2}Department of Electrical and Electronics Engineering, The Hong Kong Polytechnic University, Hong Kong.\\

    suveen.ellawela@u.nus.edu , sashenka.gamage@connect.polyu.hk , dinithid@u.nus.edu\\[0.8em]
    \url{https://github.com/SuveenE/drawdash}
}

\usepackage{bibentry}

\begin{document}

\maketitle

\begin{abstract}
\input{AnonymousSubmission/Files/0-Abstract}
\end{abstract}


\section{Introduction}
\input{AnonymousSubmission/Files/1-Introduction}

\section{Related Work}
\input{AnonymousSubmission/Files/2-RelatedWork}

\section{Methodology}
\input{AnonymousSubmission/Files/3-Methodology}

\section{System Demonstrations}
\input{AnonymousSubmission/Files/4-System_Demonstrations}

\section{Limitations and Future Work}
\input{AnonymousSubmission/Files/5-Limitations}

\bibliography{aaai2026}

\end{document}

%% file: AnonymousSubmission/Files/0-Abstract.tex
Educators frequently rely on diagrams to explain complex concepts during lectures, yet creating clear and complete visual representations in real time while simultaneously speaking can be cognitively demanding. Incomplete or unclear diagrams may hinder student comprehension, as learners must mentally reconstruct missing information while following the verbal explanation. Inspired by advances in code completion tools, we introduce DrawDash, an AI-powered whiteboard assistant that proactively completes and refines educational diagrams through multimodal understanding. DrawDash adopts a TAB-completion interaction model: it listens to spoken explanations, detects intent, and dynamically suggests refinements that can be accepted with a single keystroke. We demonstrate DrawDash across four diverse teaching scenarios—spanning topics from computer science and web development to biology. This work represents an early exploration into reducing instructors’ cognitive load and improving diagram-based pedagogy through real-time, speech-driven visual assistance, and concludes with a discussion of current limitations and directions for formal classroom evaluation.

%% file: AnonymousSubmission/Files/1-Introduction.tex
Educators frequently use whiteboards and chalkboards to draw diagrams during lectures, illustrating complex concepts through visual representations~\cite{walny2011visual, wall2005visual}. However, creating clear and comprehensive diagrams in real-time while simultaneously explaining concepts can be challenging. Incomplete or unclear diagrams may hinder student comprehension, as learners must mentally fill in missing elements or interpret ambiguous visual structures while following the spoken explanation.

Recent advances in code completion tools, such as Cursor's "TAB" completion model, have demonstrated how AI can proactively assist users by predicting and completing their intent~\cite{cursor:25}. This raises an interesting question: what if similar assistance could be provided for diagram creation in educational settings? An intelligent assistant that listens to lectures and automatically completes or improves diagrams could significantly reduce the cognitive load on instructors, allowing them to focus on teaching while ensuring students receive clear, complete visual aids.

Motivated by this vision, we developed DrawDash, an AI-powered whiteboard assistant that enhances conceptual explanations through proactive, multimodal assistance. Drawing inspiration from code completion interfaces, DrawDash follows a TAB-completion interaction model: it listens to spoken explanations, analyzes the instructor's intent, and dynamically suggests diagram refinements in real time. When the system identifies potential improvements—such as completing partial structures, adding missing components, or clarifying visual relationships—instructors can accept these suggestions with a single TAB keystroke. This seamless integration of natural language processing, visual reasoning, and human-AI collaboration aims to make diagram-based teaching substantially more effective.

We envision DrawDash as an everyday tool for diverse educational contexts, from large-scale university lectures to one-on-one tutoring sessions where someone explains a difficult concept to a peer. This paper presents an early-stage solution and describes our system architecture and design principles. Future work will include comprehensive user studies to validate the effectiveness of proactive diagram assistance in real-world teaching scenarios and to better understand how this technology can be integrated into existing educational workflows.

%% file: AnonymousSubmission/Files/2-RelatedWork.tex
Research at the intersection of Human Computer Interaction (HCI), AI, and education has recently converged on two overlapping trends that motivate DrawDash:

(1) The design and evaluation of proactive, agentic AI assistants that take initiative in human workflows.

(2) The incorporation of generative and multimodal intelligence into shared visual canvases used for design, explanation, and teaching. 
In parallel, long-standing literature on pedagogical agents and on dynamic and diagrammatic visual explanations establishes observable reasons to embed AI directly into diagram-based instruction. Below, we synthesize representative work along these themes and highlight gaps DrawDash addresses.

\subsection{Proactive, agentic assistants in interactive workflows}
HCI researchers have begun to move beyond reactive “ask-and-respond” assistants toward systems that decide when to intervene, what to propose, and how to present suggestions, reducing user effort while preserving control. A recent case study and experimental work in programming environments characterizes key design tradeoffs for proactive assistants\cite{chen:24}. This discusses how proactive assistants detect signals for intervention, how suggestions are surfaced, and how user acceptance vs. interruption tradeoffs shape productivity and trust. This line of work provides operational design rules for mixed-initiative features such as low-friction acceptance (e.g., single-key apply) and timing heuristics for proactive suggestions. 

\subsection{AI-enhanced whiteboards and multimodal canvases}

A new wave of work integrates generative AI into whiteboarding and infinite-canvas tools, revealing emergent dynamics of agency and collaboration on shared visual surfaces. Case studies of AI-enhanced whiteboarding, for example, a recent study of tldraw’s\cite{tldraw:25} generative features, show that generative affordances shift cognitive load, alter collaboration patterns, and produce novel trust and ownership concerns for users of visual canvases. 
In line with this, multimodal generative canvases, systems that combine text, voice, and visual generation to iteratively specify artifacts, demonstrate the value of iterative specification and visual dataflow affordances for complex tasks. Tools like DeckFlow illustrate how users exploit a canvas as external memory and a structured workspace for decomposition and synthesis. DrawDash is built on the same underlying philosophy, taking what the user says out loud and translates that speech into progressive visual updates on a diagram. 

\subsection{Diagrammatic instruction, progressive drawing, and multimodal learning evidence}
Research in education and learning sciences consistently shows the benefits of progressive drawing and dynamic diagrams for comprehension and retention. Scheider and team state that whiteboard-style animations, where content is revealed incrementally, tend to increase engagement and learning-relevant variables compared to static visuals~\cite{schneider:23}. Moreover, studies on cognitive and social processes of whiteboard animations, such as inserting visual information into a whiteboard animation, suggest that incremental, multimodal explanations scaffold mental model formation~\cite{krieglstein:23}. DrawDash operationalizes these principles by aligning speech segments with incremental diagram edits, thereby combining the pedagogical benefits of progressive drawing with AI-scale inference and synthesis. 

\subsection{Intelligent Autocompletion and Agentic Collaboration in Learning}
Autocompletion mechanisms, often realized through the “TAB” interaction model, exemplify how intelligent systems can anticipate and extend human intent. Traditionally used in text and programming domains, these systems enable low-friction collaboration, where users maintain control while the AI provides timely, context-aware completions. Lehmann and Buschek~\cite{lehmann:22} framed autocompletion as a general interaction paradigm for generative AI, describing how systems can interpret partial input, propose completions, and adaptively learn user preferences. Such models embody a mixed-initiative architecture, where agency is dynamically shared between human and AI collaborators.
In educational and creative contexts, similar ideas appear in intelligent tutoring and co-creative systems, where AI acts as a proactive assistant rather than a passive tool. Code assistants such as GitHub Copilot and Tabnine illustrate how predictive completion can scaffold learning, enabling novices to understand patterns and best practices through suggestion-based exploration~\cite{liu:20}.  Beyond textual or programming tasks, multimodal systems like DeckFlow\cite{croisdale:25} extend autocompletion into visual reasoning, where speech or sketches trigger intelligent diagram generation. 
Building on this trajectory, DrawDash operationalizes the TAB model within a pedagogical, diagrammatic context, where spoken explanations drive incremental visual refinement. By coupling proactive inference with low-friction acceptance, it advances a vision of agentic learning companions. These are AI systems that co-construct representations with teachers and learners, enhancing conceptual understanding through adaptive, multimodal collaboration.






%% file: AnonymousSubmission/Files/3-Methodology
Our system is implemented as a front-end/back-end architecture integrating a real-time whiteboard with generative image updates and voice-driven interaction. The major components are:

Whiteboard Canvas: We built the interactive drawing surface using the tldraw SDK (for React), which provides high-performance infinite-canvas support, shape management, undo/redo, and real-time updates. ~\cite{tldraw:25} This canvas serves as the base for both manual user sketches and AI-generated outputs.

\begin{figure}[ht]
    \centering
    \includegraphics[width=0.95\columnwidth]{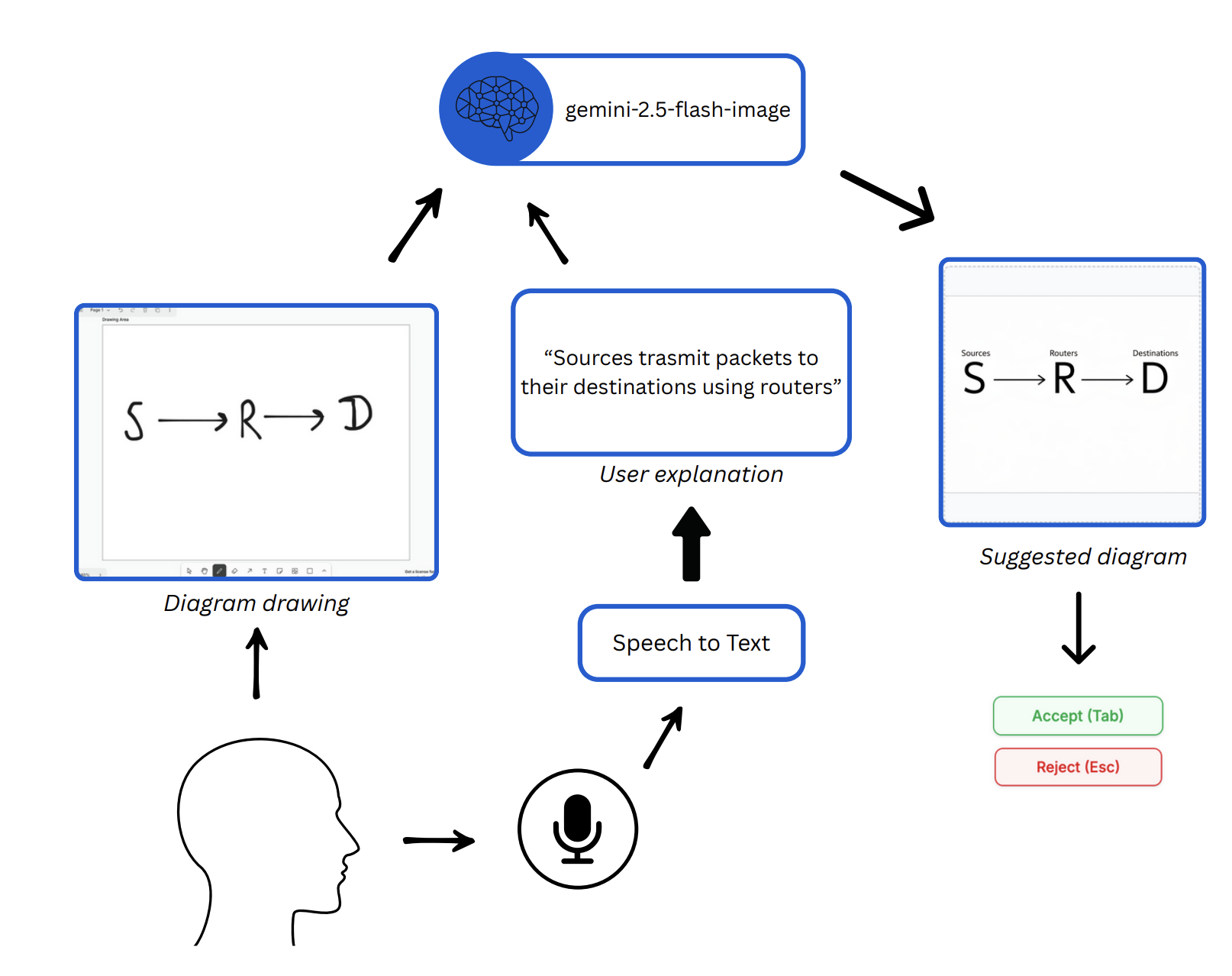}
    \caption{System architecture of DrawDash showing the integration of speech processing, visual analysis, and real-time diagram suggestion modules.}
    \label{fig:system_diagram}
\end{figure}

Proactive Diagram Generation Agent: For on-demand diagram generation and update, we use the gemini-2.5-flash-image model (aka “nano-banana”) from Gemini API via Google’s API. The model offers text-prompt and image-edit capabilities, enabling multi-turn visual refinement and consistent character/object representation across updates. ~\cite{gemini:25} We pass the user’s spoken explanation (transcribed) plus a snapshot or region of the whiteboard as input, then receive a regenerated or updated image which could be inserted back into the canvas layer. Refer Figure~\ref{fig:system_diagram} for a detailed visualization of the process.

"TAB" completion - Following the paradigm of in-editor autocompletion (“Tab”
models) like Cursor’s Tab~\cite{cursor:25} in coding tools, the user can accept the change via a single press of the "TAB" key, at which point the update merges into the canvas.

\subsection{System Workflow and Integration}

\begin{enumerate}
    \item The user begins speaking, and the speech-to-text engine produces a transcript.
    \item The system monitors for relevant keywords and update intents.
    \item Upon intent detection, the system captures the corresponding canvas region as input to the image agent.
    \item The Gemini~2.5~Flash Image model generates one or more candidate images, and the top candidate is previewed to the user.
    \item When the user taps to accept, the selected image replaces or overlays the region and becomes the new base state for further refinements.
    \item The whiteboard SDK maintains editing history, manual annotations, and layering between user-drawn and AI-generated content.
\end{enumerate}

%% file: AnonymousSubmission/Files/4-System_Demonstrations.tex


We demonstrate DrawDash's capabilities through four representative teaching scenarios that span different diagram types and educational domains. These examples demonstrate the system's ability to comprehend multimodal input (speech and partial visual diagrams) and produce contextually relevant completions.

\subsection{Methodology}

For each demonstration, we created a realistic teaching scenario consisting of:
\begin{itemize}
    \item A full or partial diagram drawn on the digital whiteboard
    \item A spoken explanation simulating natural lecture delivery
    \item An expected completion based on pedagogical best practices
\end{itemize}

The scenarios were selected to represent diverse diagram categories commonly used in educational contexts, including hierarchical structures, process flows, network topologies, and conceptual relationships. We provided the corresponding teaching scripts to a practitioner, as detailed in the test cases, and recorded their actual teaching sessions. In each case, the top figure illustrates the instructor’s initial diagram and the live transcript captured by the system, while the bottom figure shows the diagram generated by DrawDash based on that input.


\subsection{Case 1: Binary Search Tree (Computer Science)}

\textbf{Diagram Type:} Hierarchical Structure

\textbf{Scenario:} An instructor teaches binary search tree insertion, starting with a root node labeled ``50'' and verbally describing the insertion of values 30, 70, 20, and 40.

\textbf{Script:} ``We'll start with 50 as our root node. When we insert 30, since it's less than 50, it goes to the left. And 70 is greater than 50, so that goes on the right side. Now let's add 20. Twenty is less than 50, so we go left to 30, and since 20 is less than 30, it becomes the left child of 30. Finally, we'll insert 40...''

\begin{figure}[h]
    \centering
    \includegraphics[width=0.9\columnwidth]{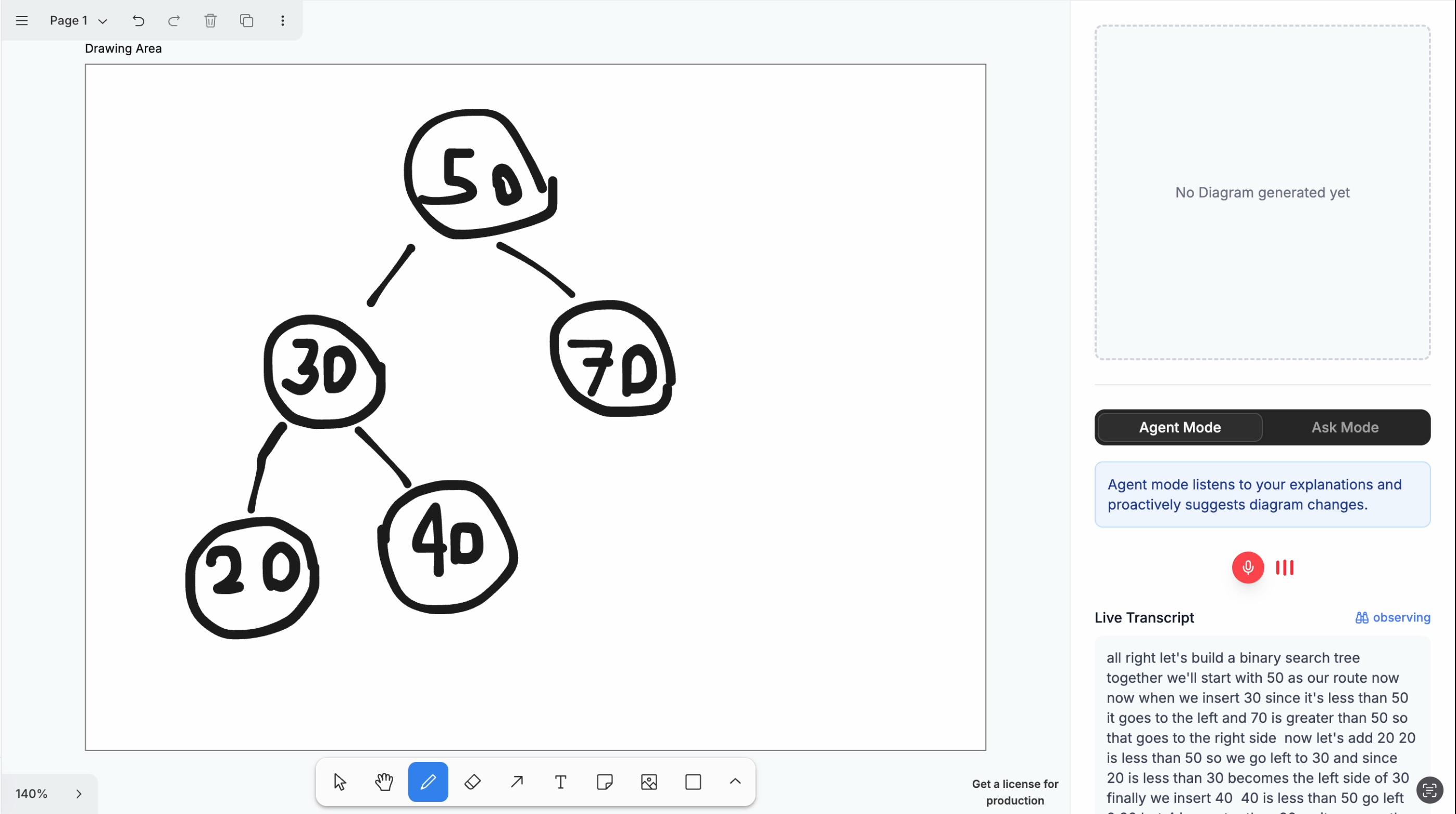}
    \vspace{0.5em}
    \includegraphics[width=0.7\columnwidth]{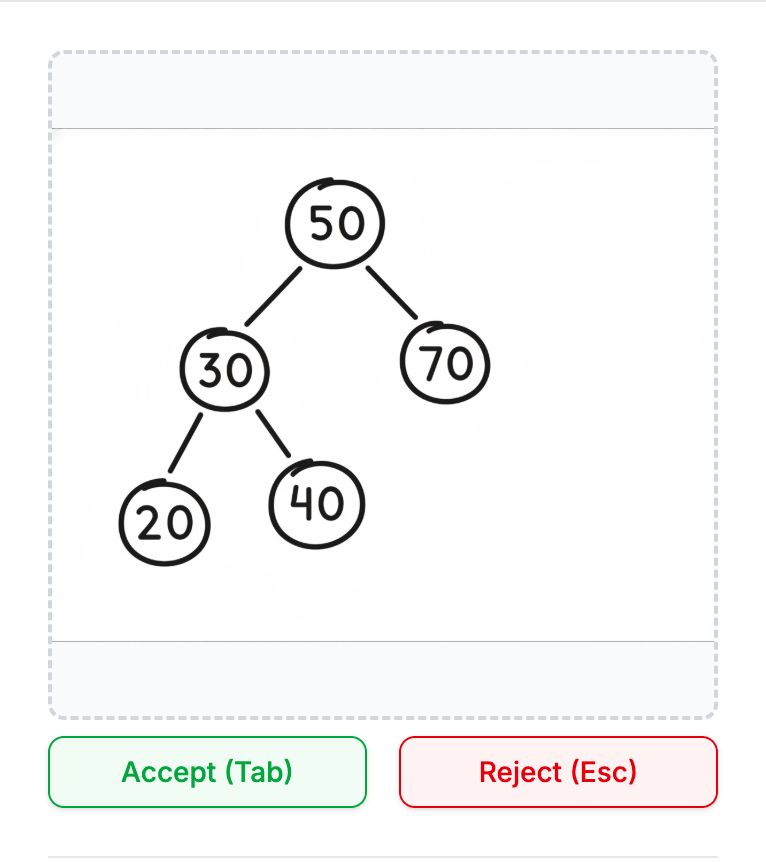}
    \caption{Binary Search Tree completion example. (\textbf{Top:}) Instructor's initial sketch showing only the root node and partial child links. (\textbf{Bottom:}) DrawDash's suggested completion illustrating the full binary search tree structure with balanced branches.}
    \label{fig:bst_demo}
\end{figure}

\textbf{System Behavior:} Refer Figure~\ref{fig:bst_demo}. DrawDash successfully identified the hierarchical structure and comparative language, refining the instructor’s rough sketch into a well-aligned binary search tree. The system adjusted node spacing, corrected edge alignment, and enhanced the legibility of the node labels, demonstrating its ability to interpret spoken relational cues and apply geometric refinements.

\textbf{Analysis:} This case demonstrates DrawDash's ability to interpret the instructor’s spoken description of relational structure and refine the corresponding visual layout. Rather than generating new content, the system focused on improving spatial alignment, node spacing, and text clarity to produce a cleaner and more readable diagram.

\subsection{Case 2: HTTP Request Flow (Web Development)}

\textbf{Diagram Type:} Process/Flow Diagram

\textbf{Scenario:} An instructor explains the HTTP request-response cycle, beginning with a box labeled ``Browser'' and an outgoing arrow.

\textbf{Script:} ``The browser first sends an HTTP request to the server. The server receives this request and processes it, maybe it needs to query a database to get some information. Once the database returns the data, the server processes it and generates an HTTP response. This response travels back to the browser...''

\begin{figure}[h]
    \centering
    \includegraphics[width=0.95\columnwidth]{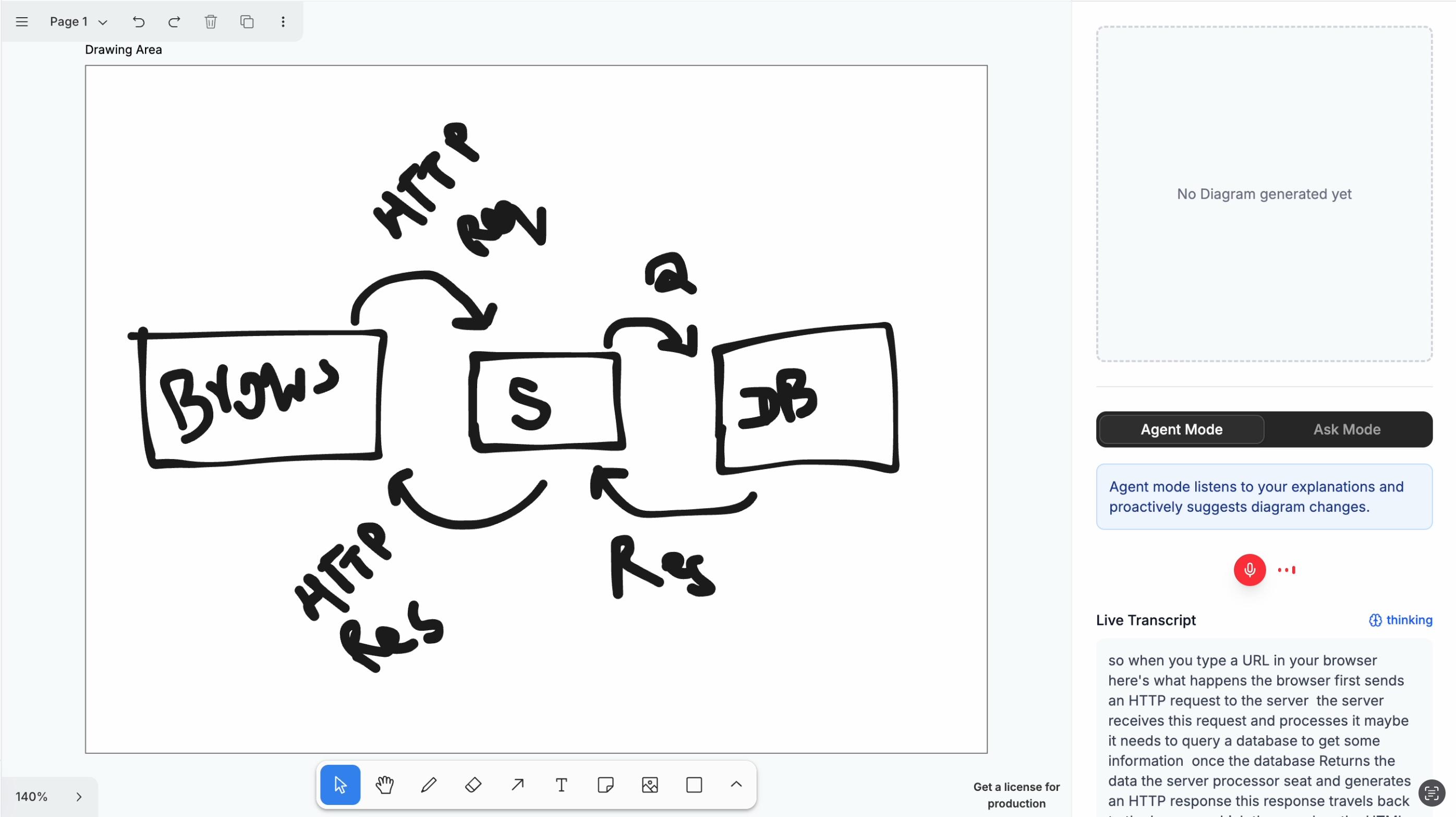}
    \vspace{0.5em}
    \includegraphics[width=0.7\columnwidth]{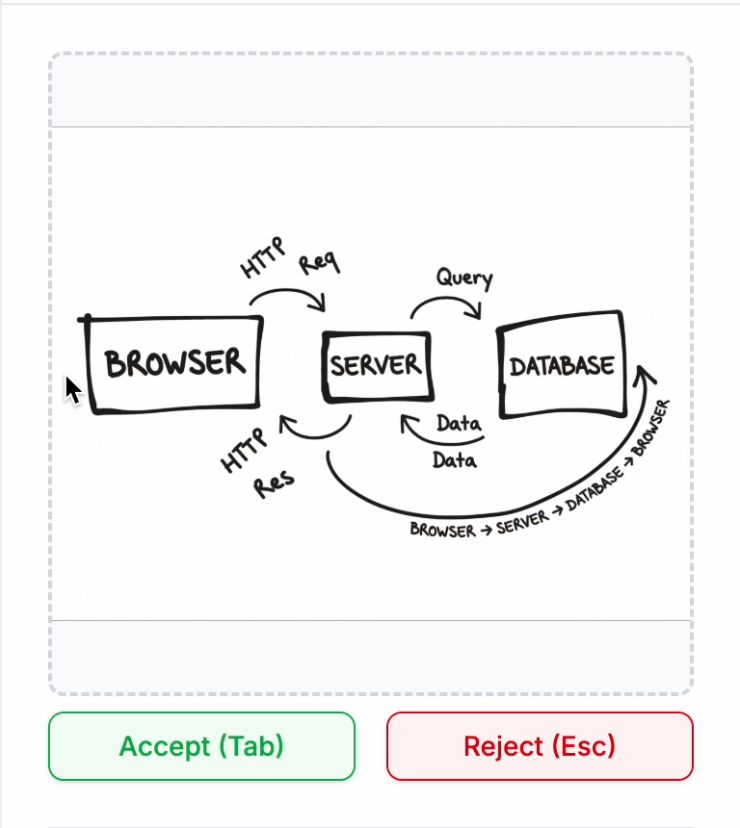}
    \caption{HTTP request flow example. (\textbf{Top:}) Instructor's initial sketch showing partial browser-server connection. (\textbf{Bottom:}) DrawDash's suggested completion illustrating the full browser–server–database interaction cycle.}
    \label{fig:http_demo}
\end{figure}

\textbf{System Behavior:} Refer Figure~\ref{fig:http_demo}. DrawDash correctly recognized the sequence of network interactions and refined the instructor’s rough sketch into a clearer process flow. It standardized the box placements for the Browser, Server, and Database, labeled the connections with directional arrows, and improved the overall visual consistency. However, the system also introduced an incorrectly placed large wraparound arrow from the Browser to the Database, which was not part of the intended flow. This error highlights a limitation in distinguishing indirect relationships from direct ones in sequential diagrams.

\textbf{Analysis:} This case demonstrates DrawDash’s ability to interpret verbal descriptions of processes and render structured, labeled diagrams. While the system successfully clarified the flow of requests and responses, the unintended browser-to-database arrow indicates the need for finer semantic control over relational inference, especially in multi-step sequences where directionality and data dependencies are hierarchical rather than direct.

\subsection{Case 3: Photosynthesis Process (Biology)}

\textbf{Diagram Type:} Conceptual/Relationship Diagram

\textbf{Scenario:} An instructor draws a simple leaf shape and explains the photosynthesis process, describing inputs and outputs.

\begin{figure}[h]
    \centering
    \includegraphics[width=0.9\columnwidth]{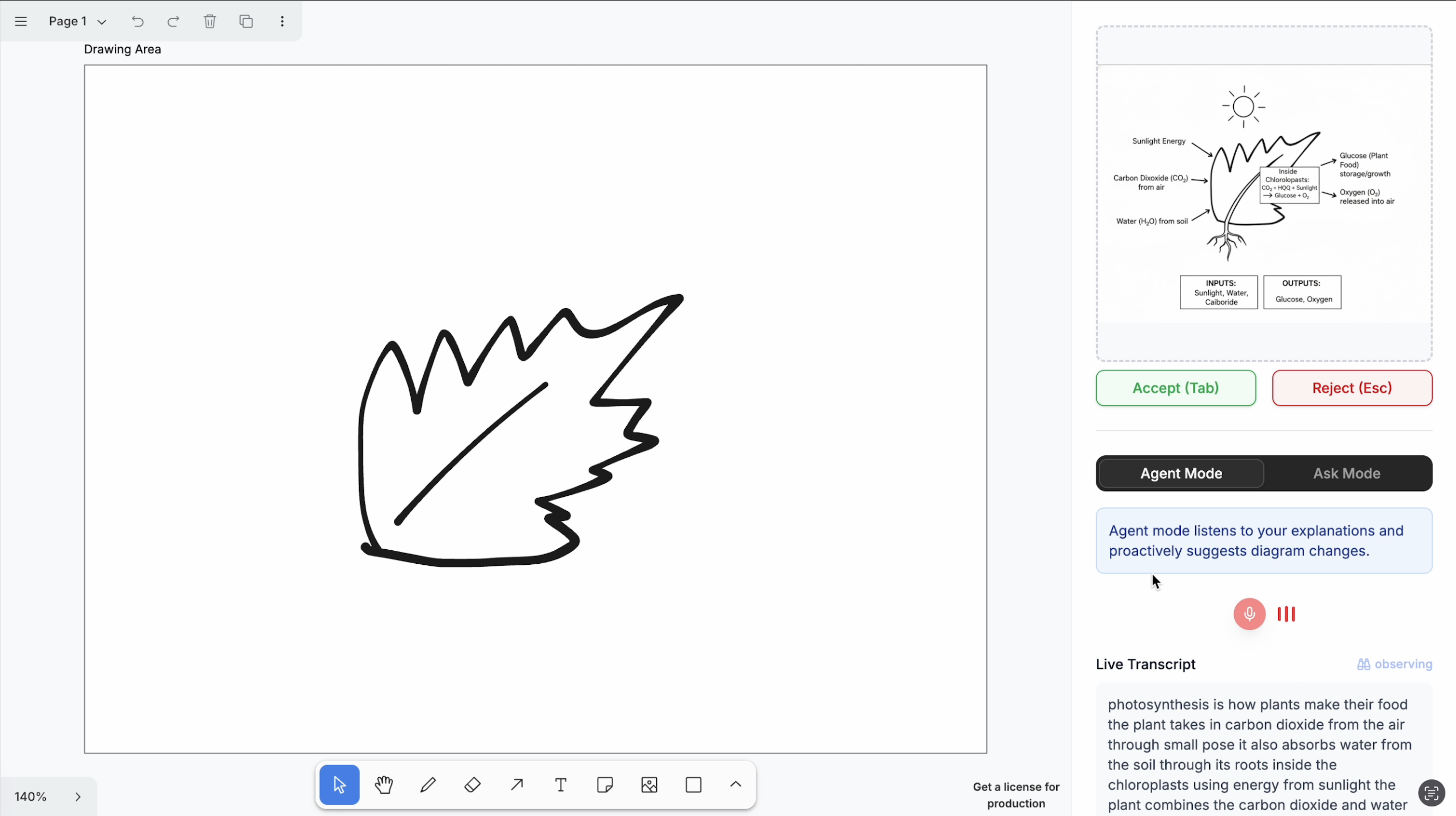}
    \vspace{0.5em}
    \includegraphics[width=0.70\columnwidth]{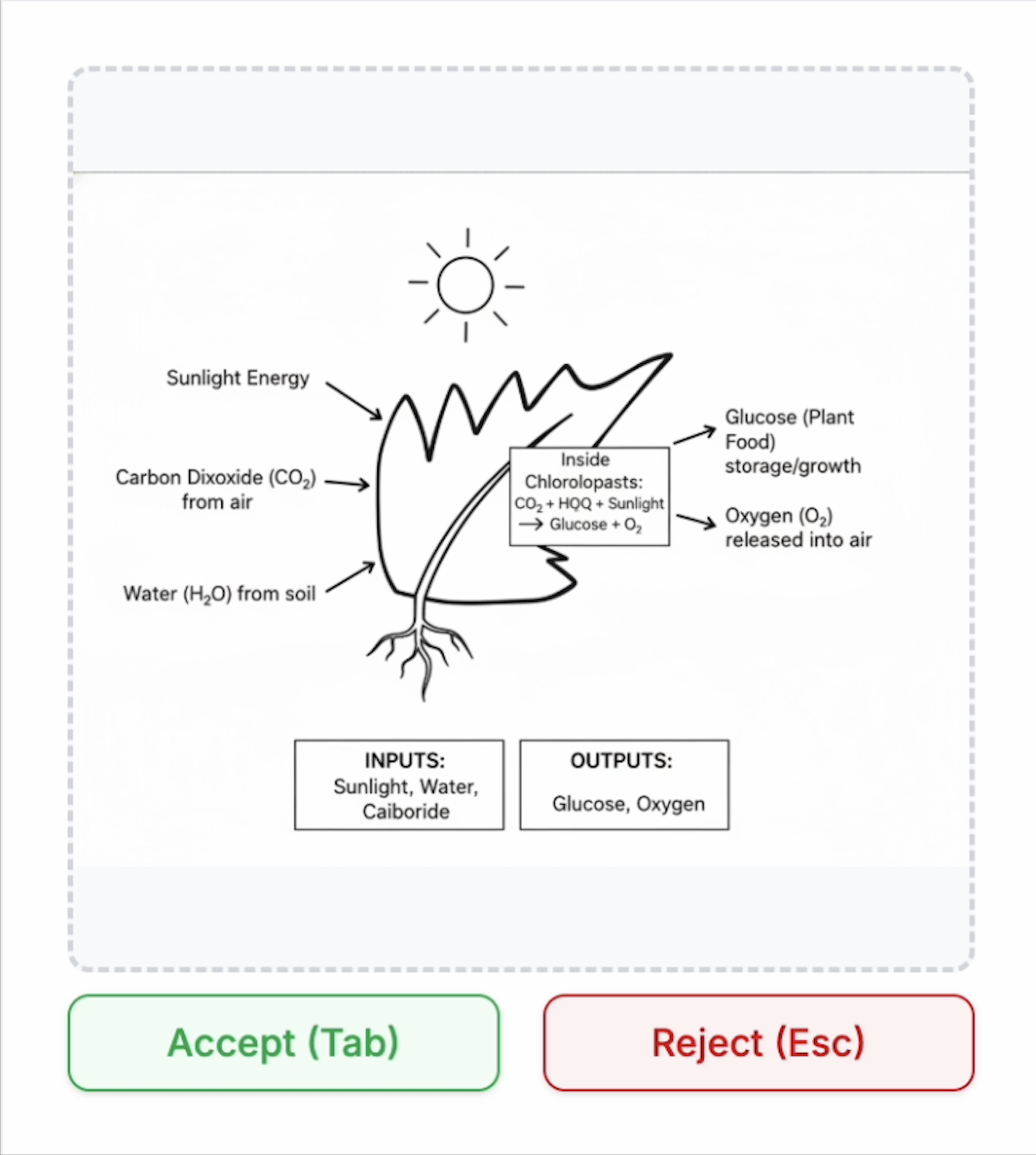}
    \caption{Photosynthesis process example. (\textbf{Top:}) Instructor's initial sketch showing a leaf with partial input labels. (\textbf{Bottom:}) DrawDash's suggested completion illustrating the full process of photosynthesis, including inputs (CO\textsubscript{2}, H\textsubscript{2}O, sunlight) and outputs (glucose, O\textsubscript{2}).}
    \label{fig:photosynthesis_demo}
\end{figure}

\textbf{System Behavior:} Refer Figure~\ref{fig:photosynthesis_demo}. DrawDash correctly identified the visual context as a leaf and expanded it into a complete photosynthesis diagram. It added directional arrows to represent input and output flows, labeled the key elements (sunlight, carbon dioxide, water, glucose, oxygen), and inserted a simple sun icon to visualize the energy source. Additionally, it generated an “Inputs” and “Outputs” summary box, enhancing the conceptual clarity of the process. The layout was clean and balanced, though some text labels slightly overlapped with arrows, indicating a need for finer spatial adjustment in automatic placement.

\textbf{Analysis:} This case demonstrates DrawDash’s ability to generalize beyond computational topics to represent natural science processes. The system successfully converted a minimal sketch and verbal explanation into a complete and pedagogically meaningful diagram. While the generated output effectively illustrated the core concept, minor spatial alignment issues suggest future work is needed for adaptive label positioning and improved handling of dense visual annotations.

\subsection{Case 4: Heating Liquid in a Beaker (Chemistry)}

\textbf{Diagram Type:} Experimental/Process Diagram

\textbf{Scenario:} The instructor sketches a beaker partially filled with liquid and inserts a thermometer into it. They then describe the setup while introducing the concept of heating using a Bunsen burner.

\textbf{Script:} “Here we have a beaker with some liquid and a thermometer placed inside. Now imagine what happens when we heat it with a Bunsen burner. As the temperature rises, the liquid starts forming bubbles—these are the early signs of boiling.”

\begin{figure}[h]
    \centering
    \includegraphics[width=0.9\columnwidth]{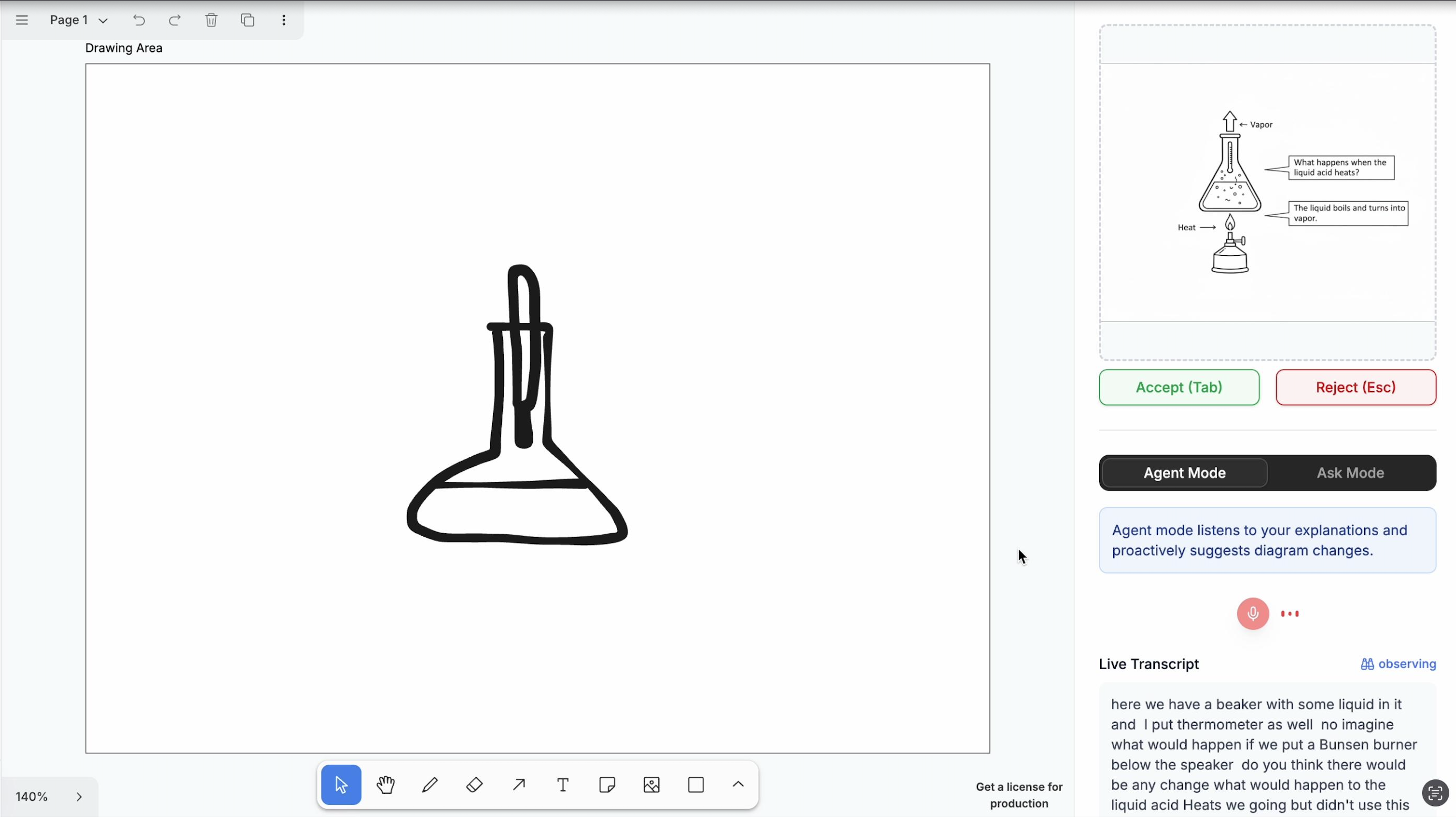}
    \vspace{0.5em}
    \includegraphics[width=0.70\columnwidth]{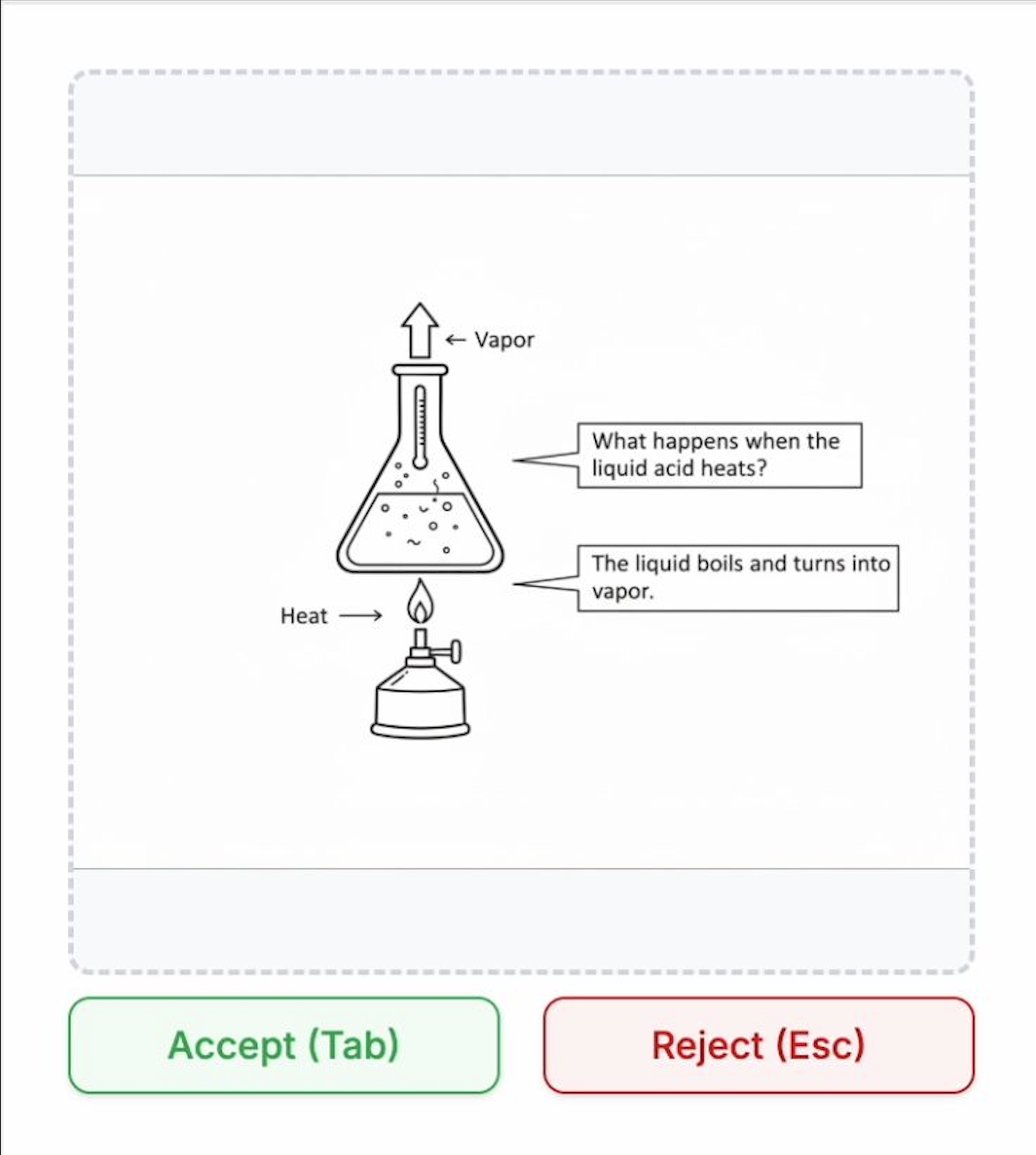}
    \caption{Heating liquid in a beaker example. (\textbf{Top:}) Instructor's initial sketch showing a beaker with liquid and a thermometer. (\textbf{Bottom:}) DrawDash's suggested completion illustrating the heating process with a Bunsen burner and rising bubbles indicating the onset of boiling.}
    \label{fig:beaker_demo}
\end{figure}

\textbf{System Behavior:} Refer Figure~\ref{fig:beaker_demo}. DrawDash successfully recognized the experimental setup and expanded the instructor’s basic sketch into a complete heating process diagram. It added a Bunsen burner icon beneath the beaker with a flame illustration, an upward arrow labeled “Vapor,” and explanatory text boxes describing the cause-and-effect relationship (“Heat” → “Liquid boils and turns into vapor”). The final output also included consistent labeling and alignment, visually clarifying the transformation process. However, the system simplified the thermometer detail from the original sketch, omitting temperature indicators, showing a limitation in preserving fine-grained drawn elements.

\textbf{Analysis:} This case demonstrates DrawDash’s capability to transform simple experimental sketches into more pedagogically complete process diagrams. By identifying causal relationships from the spoken explanation and representing them visually, the system effectively conveyed the concept of phase change. Future improvements should focus on preserving small structural details, such as measurement instruments, to maintain scientific accuracy in refined outputs.

%% file: AnonymousSubmission/Files/5-Limitations.tex
While DrawDash demonstrates promising capabilities, it remains an early-stage prototype with several limitations that guide our future directions. The system has not yet undergone formal evaluation with educators or students, so its effects on teaching efficiency, comprehension, and cognitive load remain to be empirically validated. Our current demonstrations are illustrative rather than conclusive, and controlled user studies will be essential to assess pedagogical impact and real-world usability.

Technically, DrawDash still faces challenges common to proactive multimodal systems. Its suggestion timing is based on simple heuristics—pauses in speech and visual stability—which may not align perfectly with natural teaching rhythms. Similarly, it performs best on structured diagrams such as trees and flowcharts but struggles with open-ended or highly conceptual diagrams where multiple valid completions exist. Errors in speech recognition or diagram parsing can also cascade across modalities, producing misleading or irrelevant suggestions. In addition, the system lacks long-term memory or personalization mechanisms to adapt to an instructor’s individual style and preferences.

Future work will address these limitations through comprehensive user studies and system refinements. We plan to develop adaptive timing mechanisms that learn instructor-specific pacing, integrate domain-specific knowledge for specialized diagrams, and introduce personalization based on feedback from accepted or rejected suggestions. Beyond technical improvements, we are particularly interested in exploring how proactive diagram assistance influences teaching practices, instructor autonomy, and collaboration in both classroom and remote learning contexts.

In the long term, we envision DrawDash as part of a broader ecosystem of AI-assisted educational tools—analogous to how code completion has transformed programming workflows. By reducing the mechanical effort of diagram creation, DrawDash aims to let educators focus on conceptual clarity and student engagement, ultimately democratizing access to effective visual communication in education.